\documentclass[manuscript]{aastex}
\usepackage{amsmath}
\usepackage{amssymb}
\usepackage{graphicx}
\usepackage{natbib}

\shorttitle{Vortex Creep, Toroidal Flux Lines, and Pulsar Glitches}

\shortauthors{E. G\"{u}gercino\u{g}lu \& M.A. Alpar} 

\begin{document}

\title{VORTEX CREEP AGAINST TOROIDAL FLUX LINES, CRUSTAL ENTRAINMENT, AND PULSAR GLITCHES} 

\author{Erbil G\"{u}gercino\u{g}lu}
\affil{Istanbul University, Faculty of Science, Department of
Astronomy and Space Sciences, Beyaz{\i}t, 34119, Istanbul, Turkey}

\email{egugercinoglu@gmail.com}

\and

\author{M.Ali Alpar}
\affil{Sabanc{\i} University, Faculty of Engineering and Natural Sciences, Orhanl{\i}, 34956 Istanbul, Turkey}

\email{alpar@sabanciuniv.edu}

\begin{abstract} 

A region of toroidally oriented quantized flux lines must exist in the
proton superconductor in the core of the neutron star. This region will be a site of vortex pinning and creep. Entrainment of the neutron superfluid with
the crustal lattice leads to a requirement 
of superfluid moment of inertia associated with vortex creep in excess of the available crustal moment of inertia. This will effect constraints on the equation of state. The toroidal flux region provides the moment of
inertia necessary to complement the crust superfluid with postglitch relaxation behavior fitting the observations.

\end{abstract}


\section{INTRODUCTION}

Glitches are sudden increases in rotation
rates of pulsars, with $\Delta\Omega/\Omega \sim10^{-9}-10^{-6}$,
usually accompanied by jumps in the spin-down rate, $\Delta \dot
\Omega /\dot \Omega \sim 10^{-4}-10^{-2}$ \citep{espinoza11,yu13}.
These changes tend to relax fully or partially on long timescales
(days to years), attributed to superfluid components of the
neutron star \citep{baym69}. The electromagnetic signals of pulsars
do not change at glitches, indicating that there is no change in the external torque, so that glitches reflect angular
momentum exchange between the observed crust and interior components
of the neutron star (see \citet{weltevrede11} for a notable
exception).  The  energy source of large glitches is rotational kinetic energy, which
is the minimal free energy source available for the large and
frequent exchanges of angular momentum. If additional free energy
sources like elastic or magnetic energy were involved,
the accompanying energy dissipation would exceed the observational
bounds on glitch associated thermal radiation \citep{alpar98}. 
Starquake models can account for the smaller
glitches typified by the Crab pulsar. Starquakes also act as triggers for the
large glitches \citep{alpar96}. A superfluid with quantized vortices
which can be pinned will explain the exchange of angular momentum
discontinuously as seen in the glitches \citep{packard72,anderson75},
if large numbers of vortices unpin in an avalanche which can be
self-organized \citep{melatos08}, or triggered by a starquake. 

The vortex pinning and creep model \citep{alpar84a} explains glitches and postglitch
response in terms of moments of inertia and relaxation times of the
neutron superfluid in the neutron star crust's crystal lattice,
where vortex lines can pin to nuclei. Pinning leads
to a lag $\omega=\Omega_{s}-\Omega_{c} > 0$ between superfluid and
crustal angular velocities $\Omega_{s}$ and $\Omega_{c}$. As vortex
lines pin and unpin continually by thermal activation, the lag
$\omega$ drives an average vortex current radially outward from
the rotation axis. This "vortex creep" allows the superfluid to spin down. The system evolves towards a steady state at which
superfluid and the crust spin down at the same rate,
$\dot\Omega_{s}=\dot\Omega_{c}=\dot\Omega_{\infty}$, achieved at the steady state lag $\omega_{\infty}$. In
addition to the continual spindown by vortex creep, if $\omega$
reaches a critical value $\omega_{cr}$ beyond which pinning forces
can no longer sustain the lag, a sudden discharge of the pinned
vortices occurs. The resulting angular momentum transfer to the
crust is observed as a glitch. The superfluid rotation rate
decreases by $\delta\Omega_{s}$ and the crust rotation rate
increases by $\Delta\Omega_{c}$, so that the lag decreases by
$\delta\omega=\delta\Omega_{s}+\Delta\Omega_{c}$ at the glitch. This
glitch induced change in $\omega$ offsets the creep, leading to very
slow relaxation of the spindown rate by creep as thermal activation
has a nonlinear dependence on $\omega$. There is also a linear
regime of creep leading to prompt exponential relaxation from some
parts of the superfluid. 

The superfluid core of the star is already
coupled to the crust tightly \citep{alpar84b, easson79}, on timescales short compared to the glitch rise
time, which is less than 40 seconds for the Vela pulsar \citep{dodson02}. When the interaction between vortex lines and flux lines is included the crust-core coupling timescale becomes even shorter \citep{sidery09}. The core superfluid is thus effectively included in the observed spindown of the outer
(normal matter) crust and magnetosphere. The effective crust moment
of inertia $I_c$ includes the core superfluid, so that $I_{c}\cong I$, the total moment of inertia of the star. The jump and
relaxation in the observed spindown rate of the crust indicates that the
moment of inertia fraction in crustal superfluid participating
in the glitch and postglitch relaxation is $\Delta\dot\Omega_c/
\dot\Omega_c \sim I_{\rm cr-sf}/I$. The observed $\Delta\dot\Omega_c/
\dot\Omega_c\sim10^{-3}-10^{-2}$ is consistent with the crustal
superfluid moment of inertia fraction for neutron stars. This was
proposed as a potential constraint for the equation of state
(\citet{datta93, lattimer07} and references therein). 

Superfluid neutrons in the inner crust are in Bloch states of the crust lattice. 
Their effective mass $m_{n}^{*}$ is larger than the bare
neutron mass $m_{n}$ \citep{chamel05,chamel12}. 
 This "entrainment" leaves only a fraction
of the neutron superfluid to be effectively free to store and
exchange angular momentum with the lattice
\citep{chamel06a,andersson12,chamel13}. The fractional change in the 
observed spindown rate must be
multiplied by the enhancement factor $m_{n}^{*}/m_{n}> 1$. The total moment of inertia in 
pinned superfluid sustaining vortex creep, $I_{\rm creep}$, must be large enough, such that $I_{\rm creep}/I \sim
(m_{n}^{*}/m_{n})\Delta\dot\Omega/\dot\Omega$. 
The required moment of inertia in
components of the star with pinning/creep then exceeds the
moment of inertia of the crustal superfluid, $I_{\rm creep}> I_{\rm cr-sf}$, for
reasonable neutron star equations of state \citep{andersson12,chamel13}. This suggests the involvement of the core
superfluid in glitches and postglitch relaxation. 

In the core, protons are expected to form a type II superconductor with a
dense array of flux lines \citep{baym69}. If present at all, type I superconductivity exists near the star's center, at $\rho > 2\rho_0$ \citep{jones06}. Vortices can pin to 
flux lines by minimization of condensation and magnetic energies
when vortex and flux line cores overlap \citep{sauls89, ruderman98}. Arguments for type I
superconductivity based on  putative precession \citep{link03} are invalidated by the possibility of vortex creep (see \citet{alpar05} and references therein).  
The work of  \citet{ haskell13} based on Vela glitches, concluding for either weak flux-vortex pinning or type I superconductivity also does not take creep into account. Type II superconductivity with flux-vortex pinning and creep will accommodate the observed glitch and postglitch behavior. 

The bulk of the core proton superconductor-neutron superfluid region
is likely to carry a uniform or poloidal array of flux lines. The
associated moment of inertia fraction is too large, beyond the
requirement of the entrainment effect. Furthermore, a uniform or
poloidal arrangement of the flux lines offers easy directions for
vortex line motion without pinning or creep. This will make the
effect of pinning and creep in the core dependent on the angle
between the rotation and magnetic axes, making the moment of inertia
fractions involved highly variable among different pulsars. A
toroidal arrangement of flux lines, by contrast, provides a
topologically unavoidable site for pinning and creep, and can have
conditions similar to those of pinning against nuclei in the crustal
lattice \citep{sidery09}. We discuss the toroidal arrangement of flux 
lines in neutron stars as a site of vortex pinning and creep and 
its implications for pulsar glitches.

\section{THE TOROIDAL MAGNETIC FLUX IN NEUTRON STARS}

\begin{figure}
\centering
\includegraphics*[width=1.0\linewidth]{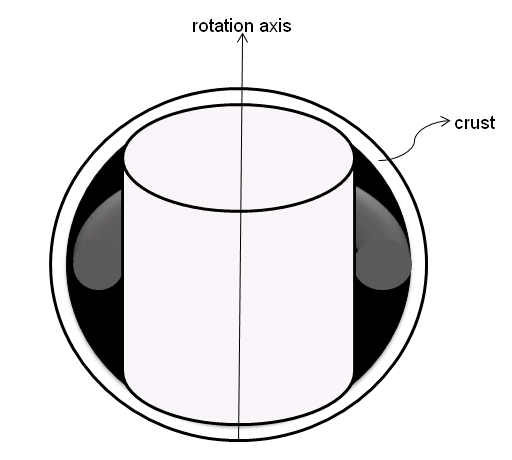}
\caption{Sketch showing the toroidal field (gray). The black shading marks the superfluid region, with moment of inertia $I_{\rm tor}$, effected by vortices creeping against the toroidal flux. For simplicity the magnetic and rotation axes are taken to be aligned.}
\label{model}
\end{figure}

In normal
(non-superconducting) stars, like the progenitors of neutron stars,
purely toroidal \citep{tayler73} or poloidal \citep{wright73}
magnetic fields are unstable. \citet{spruit99} has found that for stability of magnetic fields in stratified stars, the toroidal $B_{\phi}$ to poloidal
$B_{p}$ field ratio satisfies 
\begin{equation}
\frac{B^{2}_{\phi}}{B_{p}}< \frac{N r^{2}\rho^{1/2}}{l_{h}},
\label{criteria}
\end{equation}
where $\rho$ is density, $l_{h}$ is the horizontal
length scale of the perturbations which can be as large as the
stellar radius $R$, and $r$ is the cylindrical radial coordinate.
$N$, the buoyancy frequency of the stratified medium, has a typical
value of 500 s$^{-1}$ in neutron stars \citep{reisenegger92}. For a
very young neutron star which has not yet cooled below the
superconducting-superfluid transition temperatures, we obtain
$B_{\phi}\lesssim 10^{14}$ G by taking $l_{h}\sim r \sim R \sim10^{6}$
cm, $\rho\sim10^{14}$ g/cm$^{3}$ and $B_{p}\sim10^{12}$ G. 
 \citet{braithwaite09} has shown that stable equilibrium configurations in upper main sequence stars (neutron star progenitors) have strong toroidal fields surrounding the poloidal field. A qualitatively similar field configuration is likely to be maintained as the neutron star core
cools down and the core protons make the transition into the
type II superconductor phase. In a neutron star with a
superconducting core, a purely poloidal magnetic field in
hydromagnetic equilibrium at the crust-core boundary, though not
stable, is found to have a field strength of 10$^{14}$ G
corresponding to a surface magnetic field of
$B_{p}\sim3\times10^{12}$G, typical for radio pulsars
\citep{henriksson13}. Simulations of upper main sequence stars
\citep{braithwaite09} and neutron stars with superconducting cores
\citep{lander12,lander14} have common features. The toroidal field
component is confined within closed field lines of the poloidal
field. The poloidal field strength is maximum at the stellar center,
while toroidal field attains its largest value in the outer regions,
at $r >0.5 R$. The toroidal field is confined within the neutron
star crust for poloidal fields $\lesssim5\times10^{13}$ G \citep{lander14};
but electron differential rotation in the crust will wind the
poloidal field to generate strong toroidal flux
\citep{gourgouliatos14}, which is not likely to remain confined to
the crust, and will extend into the core. For a stable
configuration, the ratio of the toroidal and total magnetic field
energies, $E_{\rm tor}/E_{\rm mag}$ cannot be less than about 10 percent \citep{braithwaite09}. In
a model with superconducting core and proton fluid crust, this
energy ratio is found to be as large as 90 percent when crustal
toroidal fields are included \citep{ciolfi13}. Thus, simulations indicate a strong
toroidal magnetic field of 10$^{14}$ G \citep{lander14, ciolfi13},
which will be carried by the flux lines. The toroidal field is maximum at $r \sim 0.8R$, 
confined within an equatorial belt of radial extension $\sim 0.1R$
\citep{lander12,lander14}. Plausible neutron star models with relatively hard equations of state, have radii $R\cong 12$ km, insensitive to the mass in the $M\sim (1 -2 )M_\odot$ range. The density in the outer core is approximately uniform, $\rho \sim \rho_0 = 2.8 \times 10^{14}$ g/ cm$^{3}$.  The moment of
inertia fraction controlled by vortex lines passing through the
toroid, as shown in Fig.\ref{model} is estimated to be $I_{\rm tor}/I\approx 5\times 10^{-2}$. Depending on the
radial extent of the toroidal field, the moment of inertia of the
associated region can be comparable to and even larger than that of
the inner crust superfluid.

\section{POSTGLITCH RELAXATION ACCORDING TO THE VORTEX CREEP MODEL}

The observed spindown rate $\dot\Omega_{c}$ typically displays several distinct
postglitch relaxation terms with different moments of inertia and
relaxation modes, including exponentially decaying transients and
permanent changes in rotation and spindown rates.  
Depending on the pinning energy $E_{p}$ and the interior temperature $T$, vortex creep can
operate in linear or nonlinear regimes \citep{alpar89}. In the linear regime, the steady state lag $\omega_{\infty}$ is much smaller than $\omega_{cr}$. A linear creep region with moment of inertia $I_{l}$ contributes an exponentially relaxing term to the postglitch response \citep{alpar93}:
\begin{align}
\Delta\dot{\Omega}_{c}(t)=-\frac{I_{l}}{I}\frac{\delta\omega}{\tau_{l}}{\rm e}^{-t/\tau_{l}},
\label{lcreep}
\end{align}
with a relaxation time,
\begin{equation}
\label{taulin}
\tau_{l} \equiv \frac{k T }{E_{p}} \frac{R \omega_{cr}}{4 \Omega_{s} v_{0}} \exp \left( \frac{E_{p}}{kT} \right),
\end{equation}
where $v_{0}\approx 10^{7}$ cm/s is a microscopic vortex velocity. In a region where no glitch induced vortex motion takes place, $\delta\omega = \Delta\Omega_{c}$. The Vela pulsar, the
best studied glitching pulsar with glitches every $\sim 2 - 3$
years, typically exhibits three exponential transients, four
transients being resolved if the glitch is observed immediately
\citep{dodson02}. Other glitching pulsars show one or two transients
\citep{espinoza11, yu13}. 

In the nonlinear creep regime $\omega_{\infty}$ is very close to $\omega_{cr}$. The contribution of a nonlinear creep region of
moment of inertia $I_{nl}$ to the postglitch
response of the observed crust spindown rate is \citep{alpar84a}:
\begin{equation}
\Delta\dot\Omega_{c}(t)=-\frac{I_{nl}}{I}\vert\dot{\Omega}\vert \left[1-\frac{1}{1+({\rm e}^{t_0/\tau_{nl}}-1){\rm e}^{-t/\tau_{nl}}}\right],
\label{ncreep}
\end{equation}
with the nonlinear creep relaxation time
\begin{equation}
\tau_{nl}\equiv \frac{kT}{E_{p}}\frac{\omega_{cr}}{\vert\dot{\Omega}\vert}.
\label{taun}
\end{equation}
We have omitted the subscript $\infty$ from  $\vert\dot{\Omega}\vert$ as variations in the spindown rate do not exceed a few percent. 
Vortices unpinned at a glitch move through some nonlinear creep regions. These parts of the superfluid are deeply affected by the resulting sudden decrease in the superfluid rotation rate with $\delta\omega \cong \delta\Omega_{s} \gg
\Delta\Omega_{c}$. 
Creep temporarily stops, decoupling these regions from angular momentum
exchange with the crust, so that the external torque now acts on
less moment of inertia. Creep restarts after a waiting time $t_{0}
=\delta\omega/\vert\dot{\Omega}\vert$. When $t_{0}\gg \tau_{nl}$, Eq
(\ref{ncreep}) reduces to a Fermi function recovery within a time
interval of width $\sim \tau_{nl}$ around $t_{0}$. The combined
response for a distribution of waiting times $t_{0}(r) =
\delta\omega(r)/\vert\dot{\Omega}\vert$, which depends on the number
of unpinned vortices  that move through each superfluid
region, can be integrated using Eq (\ref{ncreep}). If the density of
unpinned and repinned vortices is taken to be uniform throughout some superfluid
regions of total moment of inertia $I_A$, representing a mean field approach, then the integrated contribution to $\Delta\dot\Omega_{c}(t)$ 
is characterized by a constant second derivative $\ddot\Omega_{c}$ 
with which $\dot\Omega_{c}(t)$ recovers its preglitch value after a waiting 
time $t_0$ corresponding to the maximum initial postglitch offset 
$\delta\omega$ in these unpinning-repinning regions \citep{alpar84a}. 
When initial transients are over, this slower response takes over. This behavior 
prevails in the interglitch timing of the Vela pulsar, and its healing signals
the return to preglitch conditions, providing an estimate of the
time of occurrence for the next glitch. Such constant
$\ddot\Omega_{c}$ is common in older pulsars \citep{yu13}, and scales
with the parameters of the vortex creep model \citep{alpar06}. Part
of the glitch in $\Omega_{c}$, associated with moment of inertia
$I_B$, never relaxes back. This corresponds to vortex free regions
$B$ interspersed with the unpinning-repinning creep regions $A$. The
vortex free regions $B$ are analogous to capacitors in a circuit:
they do not support continuous vortex currents and do not contribute
to the spindown, transferring angular momentum only at glitches
when the unpinned vortices pass through. The glitch magnitude is
given by the angular momentum balance \citep{alpar93}
\begin{equation}
I_{c}\Delta\Omega_{c}=(I_{A}/2+I_{B})\delta\Omega_{s}.
\label{magnitude}
\end{equation}

\section{VORTEX CREEP AGAINST TOROIDAL FLUX LINES}

Junctions with toroidal flux lines inevitably constrain motion of the vortex lines.
Entrainment of the neutron and proton mass currents in the core
endows a vortex with a magnetic field of $B_v=[(m_p-m_{p}^*)/m_n
[\Phi_0/\pi\Lambda_*^2]\sim10^{14}$ G, while the magnetic field in a
flux line is $B_\Phi=[\Phi_0/\pi\Lambda_*^2]\sim10^{15}$ G
\citep{alpar84b}. The pinning energy due to magnetic interaction
between a vortex and a flux line is of the order of $E_p = (B_{v}
B_{\Phi}/4\pi) \times V$, where $V\cong 2\pi\Lambda_{*}^{3}$ is the
overlap volume with the interaction range given by the London length
$\Lambda_*=29.5[(m_p^*/m_p){x_p}^{-1}\rho_{14}^{-1}]^{1/2}$ fm
\citep{alpar84b}. In the above expressions $\Phi_0\equiv hc/2e$ is
flux quantum, $m_{p}^{*}$ and $m_{p}$ are effective and bare mass of
the proton, $x_{p}\sim 0.05$ is the proton fraction in
the outer core and $\rho_{14}$ is density in units of 10$^{14}$
g/cm$^{3}$. \citet{chamel06b} find $m_{p}^{*}/m_{p}\sim0.5-0.9$, with
$m_{p}^{*}/m_{p}\cong 0.5$ indicated by limits on crust-core coupling
from the resolution of Vela glitches \citep{dodson02}. A rough
estimate gives $E_{p}\sim 6$ MeV, though there is a wide range of
estimates $E_{p}\sim 0.1-10$ MeV \citep{sauls89, chau92}. Taking the
range of the pinning force as $\sim \Lambda_*$ and the average
length between junctions as the spacing between flux lines,
$l_\Phi=(B_{\phi}/\Phi_0)^{-1/2}$, the maximum lag $\omega_{cr}$ that can
be sustained by pinning forces is given by the Magnus equation $
\rho\kappa R\omega_{cr}= E_p/ l_\Phi\Lambda_* $. The temperature at
the crust-core boundary can be estimated for 
cooling via the modified Urca process \citep{yakovlev11}, or by
relating the inner crust temperature to surface temperature
measurements \citep{gudmundsson83}. Both methods give interior
temperatures of $10^{8}-10^{9}$ K. With these ranges of $E_{p}$ and
$kT$, vortex creep will be in the
nonlinear regime. The nonlinear creep relaxation time does not have
the uncertainties of the $E_p$ estimate when divided by
$\omega_{cr}$, giving, scaling with Vela pulsar parameters,
\begin{align}
\tau&\simeq60\left(\frac{\vert\dot{\Omega}\vert}{10^{-10}\mbox{
rad/s$^{2}$}}\right)^{-1}\left(\frac{T}{10^{8}\mbox{
K}}\right)\left(\frac{R}{10^{6}\mbox{ cm}}\right)^{-1}
x_{p}^{1/2}\times\nonumber\\&\left(\frac{m_{p}^*}{m_{p}}\right)^{-1/2}\left(\frac{\rho}{10^{14}\mbox{ gr/cm$^{3}$}}\right)^{-1/2}\left(\frac{B_{\phi}}{10^{14}\mbox{ G}}\right)^{1/2}\mbox{days,}
\label{tau}
\end{align}
with $\rho = 2 \times 10^{14}$ g/cm$^{3}$ and $x_{p}= 0.05$ we
obtain $\tau \cong 30$ days. The toroidal flux line region has
no obvious structures to provide vortex traps. The crust lattice with its domains and dislocations, 
can provide vortex trap regions $A$ and vortex free
regions $B$ interspersed with them, and is the locus of crust
breaking to trigger vortex unpinning. Thus it is likely
that vortices are unpinned from traps in the crust superfluid. As these vortices move outwards, they do not traverse the toroidal flux
region which lies further in. There is therefore no change in the superfluid rotation rate in the toroidal flux
region. The offset time here is determined by the glitch in the
observed rotation rate of the crust:
\begin{equation}
t_0=\frac{\Delta\Omega_c}{\vert\dot{\Omega}\vert}=7\left(\frac{t_{sd}}{10^4\mbox{
yr}}\right)
\left(\frac{\Delta\Omega_c/\Omega_c}{10^{-6}}\right)\mbox{days},
\label{offset}
\end{equation}
where $t_{sd}=\Omega/2\vert\dot{\Omega}\vert$ is pulsar spindown age. Expanding Eq.(\ref{ncreep}) in $t_{0}/\tau < 1$, we obtain
\begin{align}
\Delta\dot{\Omega}_{c}(t)=-\vert\dot{\Omega}\vert\frac{I_{\rm tor}}{I}\frac{t_{0}}{\tau}{\rm e}^{-t/\tau}.
\end{align}

We omit the mass entrainment correction $m_{p}^{*}/m_{p} < 1 $ in
the core superfluid. Its effect on estimating the moment of inertia
of the superfluid controlled by the toroidal field region will be
within the uncertainties in the actual extent of
the toroidal region. Taking into account
$m_{p}^{*}/m_{p} < 1 $ will decrease rather than increase the value
of $I_{\rm tor}$ to be inferred from $\Delta\dot{\Omega}_{c}$. This
response of the nonlinear creep against toroidal flux lines is of
the same form as the linear creep response of inner crust superfluid
associated with postglitch exponential relaxation,
Eq.(\ref{lcreep}), but with the nonlinear relaxation time and offset
time given by Eqs.(\ref{tau}) and (\ref{offset}).

\section{CONCLUSIONS}

The entrainment effect for the crustal superfluid
requires more moment of inertia in extra-crustal superfluid
regions with pinning and creep in order to account for the observed glitch related changes in the spindown rates of pulsars. The toroidal configuration of flux
lines in the outer core can provide the site for this. Creep response in this region provides an exponentially relaxing contribution to the glitch in the spindown rate. 
For Vela \citep{alpar93} and Crab
\citep{alpar96} glitches, the crustal superfluid with exponential
relaxation makes up the largest part of the moment of inertia involved, $\sim
10^{-2}I$, without taking entrainment into account. 
There is a particular exponential relaxation component with
$\tau \cong 32.7$ days in agreement with our estimate for the toroidal
flux line region for the first nine Vela glitches. The amplitudes of
this exponential relaxation are in the range
$\Delta\dot{\Omega}_{l}\cong(0.58-1.21)10^{-2}\dot\Omega$
\citep{chau93}. The nonlinear creep response of the toroidal flux
line region, as well as the linear creep response of crustal
superfluid employed in earlier work can contribute to the observed
$\Delta\dot{\Omega}_{l}$, as both components relax with similar
relaxation times and commensurate moments of inertia. Taking into
account vortex creep against toroidal flux lines, the moment of
inertia fraction $I_l/I$ in the crustal superfluid involved in
exponential relaxation leads to a new constraint on the total
crystalline crust moment of inertia $I_{\rm crust}$
\begin{equation}
\frac{I_l}{I} =\left(\frac{\Delta\dot{\Omega}_{l}}{\dot\Omega} -\frac{I_{\rm tor}}{I}\right) \frac{m_{n}^{*}}{m_{n}} \sim 10^{-3}- 10^{-2}< \frac{I_{\rm crust}}{I},
\end{equation}
which in principle can lead to constraints on the equation of state
\citep{lattimer07}, if uncertainties in $I_{tor}/I$, $m_{n}^{*}/m_{n}$ and
the location of the crust-core boundary are resolved.  With
entrainment in the crustal superfluid, the angular momentum balance,
Eq.(\ref{magnitude}), becomes
\begin{equation}
\frac{\Delta\Omega_{c}}{\delta\Omega_{s}}=\frac{m_{n}}{m_{n}^{*}}\frac{I_{A}/2+I_{B}}{I_{c}}\lesssim\frac{m_{n}}{m_{n}^{*}}\frac{I_{\rm cr-sf}}{I-I_{\rm cr-sf}-I_{\rm tor}}.
\end{equation}

Using the
analysis of Vela pulsar glitches with the vortex creep model
\citep{alpar93,chau93} and estimate of $I_{\rm cr-sf}/I\simeq
4\times10^{-2}$ \citep{lattimer07}, we obtain $m_{n}^{*}/m_{n}\lesssim
2.2-4$. This range accounts for a  density range 
$\rho \gtrsim 6.4\times10^{13}$ g/cm$^{3}$ in the inner crust 
\citep{chamel12,chamel13}. It should be noted that calculations of the 
enhancement factor assume a bcc lattice that may not be valid 
\citep{kobyakov13}; uncertainties about defects and impurities as well 
as "pasta" structures may also lead
to smaller enhancement factors \citep{chamel13}. Recent work explores if plausible neutron star equations of state allow for a thicker crust to accommodate large enhancement factors  \citep{steiner14, piekarewicz14}. 

The magnetar 1RXS J170849.0-4000910 \citep{kaspi03} and the radio
pulsar PSR B2334+61 \citep{yuan10} underwent glitches with
exponential relaxation for both of which
$\Delta\dot\Omega_{c}/\dot\Omega_{c}\sim 0.1$, indicating moments of
inertia larger than the crustal superfluid even without entrainment.
The response of the toroidal field region can account for these as
well:  In regions without glitch associated vortex motion the
response would still be exponential relaxation, and the toroidal
flux line region would contribute a similar response, providing the
extra moment of inertia. In older pulsars, the linear creep regions
of the crustal superfluid progressively become nonlinear creep
regions, and relaxation times calculated by Eq.(\ref{tau}) become longer. 
In this case glitches would be step like
increases with no significant relaxation. Such behavior is indeed
observed \citep{espinoza11,yu13}. The
exponential relaxation time $\tau$ in Eq.(\ref{tau}), if identified
from pulsars of different ages as corresponding to the toroidal flux
region, can yield information on microphysical parameters and the
location of the crust-core boundary.

We have given a proof of principle about the role of vortex pinning
and creep response from the toroidal flux region in the outer core
of the neutron star. The superfluid controlled
by pinning and creep in this region can complement the crust superfluid to 
accommodate the moment of inertia requirements of entrainment. The vortex
creep relaxation times are consistent with analysis of
postglitch response in the Vela glitches and scaling of the model to
other pulsars.

\acknowledgments We thank the referee for constructive comments. This work is supported by the Scientific and Technological Research
Council of Turkey (T\"{U}B\.{I}TAK) under the grant 113F354. M.A.A.
is a member of the Science Academy (Bilim Akademisi), Turkey. 


\end{document}